# Whispering-Gallery Mode Resonator Technique with Microfluidic Channel for Permittivity Measurement of Liquids

Alexey I. Gubin, Alexander A. Barannik, Nickolay T. Cherpak, Irina A Protsenko, Sergey Pud, Andreas Offenhaeusser, Svetlana A. Vitusevich

*Abstract*— Studies of biochemical liquids require precise determination of their complex permittivity. We developed a microwave characterization technique on the basis of a high-quality whispering-gallery mode (WGM) sapphire resonator with a microfluidic channel filled with the liquid under test. A novel approach allows obtaining the complex permittivity of biochemical liquids of sub-microliter/nanoliter volumes with high accuracy. The method is based on special procedure of analysis of the interaction of electromagnetic field with liquid in WGM microfluidic resonator and measurements of both the WGM resonance frequency shift and the change of the inverse quality factor. The approach is successfully applied to obtain the complex permittivity in the Ka band of glucose, albumin bovine serum, lactalbumin and cytochrome C aqueous solutions using the developed microwave technique.

## I. Introduction

Accurate electromagnetic characterization is of crucial importance for a wide range of substances [1]. However, the challenge becomes more complex for evaluating the electromagnetic parameters when small amounts of substances are to be tested, especially in the case of biochemical liquids [2]. In this respect, novel approaches for supplying, positioning and testing the liquid have to be developed [3,4,5].

It is well known that there are methods and technical approaches to study the properties of condensed matter using both resonator and non-resonant electrodynamic structures. Each of the approaches has its advantages and disadvantages. Resonant approaches are the most accurate and sensitive techniques for microwave characterization of substances including founding complex permittivity of liquids. However resonator technique allows measuring only at one or several resonant frequencies with a single device [1].

Waveguides or planar transmission lines allow performing measurements in a large bandwidth with frequency independent characteristic impedance and propagation constant. This peculiarity makes them useful in the broadband study of substances depending on frequency [6,7,8]. On other hand their measurement accuracy and sensitivity are not precise enough. With a broadband sensor the most promising frequency range for a specific substance can be found, whereas a resonant based sensor can be used to extract detailed information on this substance at this particular frequency.

Whispering-gallery mode (WGM) resonator techniques provide advanced properties comprising open resonator system, high quality factor, accessibility of rigorous analysis of electromagnetic field distribution in the resonator, controllability of mode coupling with feeder lines. Using microfluidic channels [3,4,5,6,8,9] in the microwave band allows the contactless measurement of different liquids that is important for various biochemical applications. The technique based on a WGM resonator covered by a plastic layer with a microfluidic channel [10] has additional advantages compared to other WGM techniques such as a closed measurement cell system and the utilization of a constant small amount (nL range) of the liquid under test (LUT). These advantages lead to increased accuracy and high sensitivity of the technique based on WGM resonator with the microfluidic channel.

The complex permittivity of the liquid under test can be obtained by analytical (or numerical) solutions in the case of an open coaxial line technique after the setup calibration procedure [11]. On the other hand, this method demands relatively large amounts of liquids (mL range). At the same time, the WGM resonator technique allows the amount of the LUT to be reduced. However, the analytical solution of the electromagnetic problem for the WGM resonator with the microfluidic channel has not been reported yet. Therefore obtaining the absolute values of the complex permittivity from the measurement data was not possible. In [12] it was shown that the numerical methods may be used to obtain the permittivities of LUTs. The permittivity of cytochrome C aqueous solution filling the microfluidic channel was obtained by fitting of modeling data to the experimental ones. However fitting procedure takes too much time and using of such a

A. I. Gubin, A. A. Barannik, N. T. Cherpak, I. A Protsenko are with the Department of solid-state radiophysics, O. Usikov Institute for Radiophysics and Electronics NAS of Ukraine, 61085 Kharkiv, Ukraine (e-mail: alex_gubin@mail15.com).

S. Pud, A. Offenhäusser, S. A. Vitusevich are with Peter Grünberg Institute (PGI-8), Forschungszentrum Jülich, 52425 Jülich, Germany (Corresponding author e-mail: s.vitusevich@fz-juelich.de).

S.Pud gratefully acknowledges a research grant from the German Academic Exchange Service (DAAD).

procedure is reasonable for a couple of points only.

Here we present the results demonstrating a novel approach for the precise estimation of LUT complex permittivity by utilizing a WGM resonator with a microfluidic channel. The solution of the electromagnetic problem for the WGM resonator was found with assistance of numerical calculations using modeling software (COMSOL Multiphysics) and comparison of calculated and experimentally obtained results (section III.D). The proposed procedure was successfully verified by test measurements of a known biological liquid: an aqueous solution of glucose (section IV.A). The complex permittivity of glucose, albumin bovine serum, lactalbumin and cytochrome C aqueous solutions of various concentrations was obtained on the basis of measured data of the WGM resonator technique (section IV.B).

## II. EXPERIMENTAL AND MODELING DETAILS

### A. Experimental Technique

We used the technique to investigate small volumes of liquids, which were put into the measurement cell developed on the basis of WGM dielectric resonator covered by plastic layers with a microfluidic channel filled with LUT [13].

The measurement setup also consisted of two sapphire dielectric waveguides as input and output transmission lines and the positioning system with temperature sensor and Peltier element (Fig.1). Such a measurement cell is placed in the temperature isolated box. Resonant frequency and quality factor were measured using 8722C Network Analyzer. The setup has a temperature stabilization system which allows the temperature of the resonator and LUT to be stabilized with an accuracy of better than ± 0.01°C.

Sapphire disk resonator (d=14.5 mm, h=2.5 mm) is used in our measurement cell. The temperature variation of the resonator quality factor is negligible and the variation of resonant frequency shift of such a resonator is about 2 MHz per Celsius degree. The error of the resonant frequency determination caused by resonator temperature variation is lower than 20 kHz in our case. Such a measurement cell allows characterizing the resonator with LUT at a number of resonant frequencies (on different modes) with spacing about 1.8 GHz for the frequency range of 27-40 GHz. However the measurement on different modes usually requires changing the resonator coupling with the dielectric waveguides. All the measurements in our work were performed on a single mode (on a single resonant frequency) to keep identical all measurement conditions.

The measurement cell was optimized using several approaches. Different plastics were investigated to choose the optimal material for making the microfluidic system. Zeonor 1420R was used as suitable material for our measurements. It has a relatively low dielectric loss (inverse quality factor of the resonator covered with such a plastic is about $6.25 \times 10^{-6}$) and can be processed using thermo-compression in a form of the microfluidic system with improved quality of the surface in the measurement cell.

The designed microfluidic systems (see Fig.1) were used to perform investigation of liquids. The channel shape is a cylindrical tube at mid-height of the 0.69 mm thick plastic layer. The channel edges are connected hermetically to the thin metal tubes. This allows us to investigate different liquids including water and bio liquids without a problem related to the consideration of the liquid densities. A position of the channel was optimized according to simulated redistribution of electromagnetic field using COMSOL Multiphysics in the case of water filled microfluidic channel of 0.2 mm inner diameter. The abovementioned optimizations enable high sensitivity of measurements of bio-liquids since most of them are aqueous solutions.

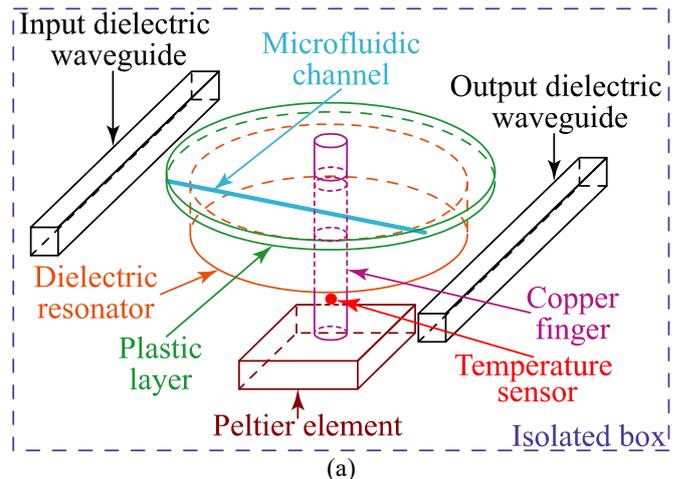

(a)

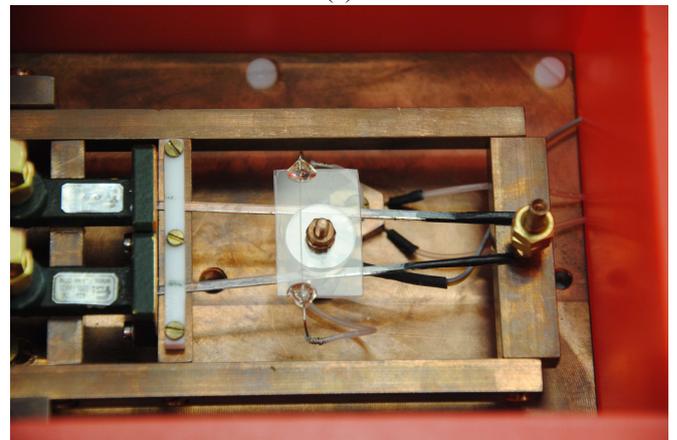

(b)

Fig. 1. Measurement cell consisting of WGM resonator with a plastic layer and a microfluidic channel, a pair of dielectric waveguides with coaxial-to-waveguide transitions in a thermally stabilized box: (a) – schematic, (b) - photo.

### B. Numerical modeling

Different commercial software products were tested for the simulation of measurement cells based on a WGM disk dielectric resonator. A number of WGM dielectric resonators were simulated and the data were compared with analytical calculations [14]. COMSOL Multiphysics was chosen as the most appropriate and accurate software for solving our problem (Fig. 2). The calculation model is based on a consideration of the sapphire ($\varepsilon_\perp = 9.4$, $\varepsilon_\parallel = 11.59$, $\tan \delta = 2.5 \cdot 10^{-5}$) disk resonator and plastic ($\varepsilon = 2.38$, tan δ = 8.7·10⁻⁴) layer with the microfluidic channel. The plastic and sapphire loss tangent values were obtained experimentally. The dimensions are used as in the experiment [15] (see the section II.A). The model is realized in the Radio Frequency module using the Eigenfrequency solver. To discretize the model all model parts are divided into the tetrahedral elements. The complete mesh consists of 84440 elements.

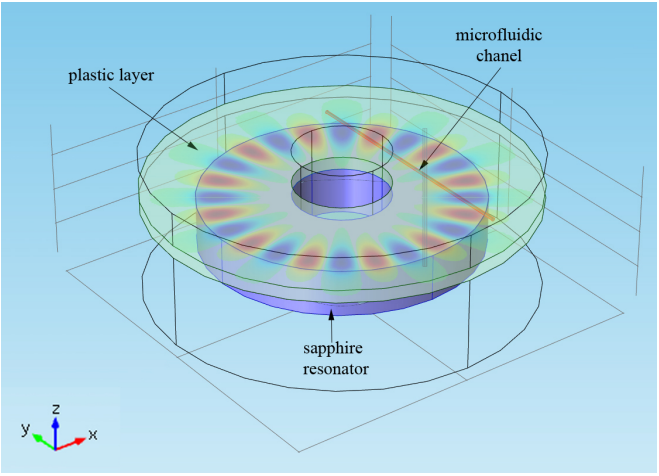

Fig. 2. Numerical geometric model for simulation of the measurement cell and axial component of electric field distribution.

The $HE_{12\,1\,1}$ mode used for measurements is twice degenerated on the azimuthal index in the resonator without a microfluidic. In a resonator with a water-filled microfluidic channel on a top, degeneration is removed and the $HE_{12\,1\,1}$ mode splits. The first split mode $HE_{12\,1\,1}$ has a resonant frequency of 35.39268 GHz with a quality factor $Q = 3500$ and the second one has a resonant frequency of 35.39294 GHz with $Q = 3930$. In the experiment, it is not possible to distinguish these modes because the difference between the resonant frequencies is too small. Therefore the results of calculations for the two modes were averaged to obtain a single peak as in the experiment. The calculated data were then analyzed together with the data obtained experimentally. A special calibration procedure was introduced to obtain accurate geometric parameters of the real microfluidic channel filled with LUT and its position in the plastic layer. The obtained data were used to find the absolute value of the complex permittivity of glucose solutions and other biological aqueous solutions as is shown below.

### III. Special Calibration Procedure

To obtain the absolute value of the complex permittivity of LUT from the measured data, we performed a special calibration of the experimental setup by simulating it using COMSOL Multiphysics.

#### A. Fitting Procedure for Precise Identification of Microfluidic Channel Diameter and Position

In order to precisely determine the interaction of the liquid inside the microfluidic channel with the electromagnetic field, the geometry of the system should be determined with high precision. The measured dimensions of the microfluidic channel and its position in the plastic layer can be accurately refined by fitting the experimental data with calculated data using COMSOL Multiphysics for our case of a WGM resonator with a microfluidic channel. The series of measurements: both the WGM resonance frequency shift and the change of the inverse quality factor were performed with different positions of the plastic layer with a microfluidic channel filled with distilled water and a channel without liquid (air inside the microfluidic channel). The measurements with air- and water-filled microfluidic channel were used to determine the real diameter. In the calculations, the known permittivity of water was used as a well-known reference [16].

We used the refinement procedure described below. In the case of a real sample, the position of the microfluidic channel in the plastic layer is displaced from the middle of the layer in the axial (z) direction due to the fabrication procedure. In order to obtain the value of such a shift, we fitted the calculated data to the measured results for a microfluidic channel filled with water. The measurements were performed in two positions of the plastic layer with the microfluidic channel when it was facing the resonator with its lower and with its top part. In such geometries, we can evaluate the precise channel displacement. This means that the distance of the microfluidic channel from the resonator surface is half the plastic thickness plus the value of the shift in one case and half of the plastic thickness minus the value of the shift in the other. In order to fit the experimental measurement, we considered two positions of the channel, corresponding to the two cases mentioned above. The position of the channel was changed in the same way in the simulation model in order to fit the simulated pair of results to the measured results. The refined position of the microfluidic channel in the radial (x) direction as well as the refined channel radius were obtained from fitting the calculated data to measured results for a microfluidic channel filled with air and water.

The geometrical dimensions in the process of fitting were varied only within the measurement accuracy of the channel dimensions. The dimensions were varied in the calculations to achieve the best fitting of the simulated data to the measured data. The geometrical dimensions as well as the position of the microfluidic channel obtained from the best fitting were found to be as follows: diameter of the microfluidic channel is 0.188 mm; shift of the microfluidic channel from the middle of the plastic layer depth in the direction of the resonator is 0.0325 mm and in a radial direction it is 4.1 mm from the resonator center.

#### B. Reliability of the method

Measurements of resonant frequency shift $f_{liq} - f_{air}$ of the resonator with microfluidic channel filled with liquid in respect to air-filled one and losses in liquid, which can be found as a difference in inverse quality factors of the resonator with liquid filled microfluidic channel in respect to air-filled one $1/Q_{liquid} - 1/Q_{air}$ for different well known liquids, were performed to check reliability of the numerical calculations. Methanol, ethanol, propanol, acetone and water were chosen as test liquids with well-known values of the permittivity [16]. The temperature of the resonator and LUT was stabilized at the 25°C. The measurement was performed on HE$_{12\,1\,1}$ mode (frequency around 35.5 GHz). The frequency shift $f_{liq} - f_{air}$ and changes of inverse quality factors $1/Q_{liquid} - 1/Q_{air}$ of the resonator with liquid filled channel in respect to air as a function of the real and imaginary parts of LUT permittivity are shown in Fig. 3a and Fig. 3b, respectively.

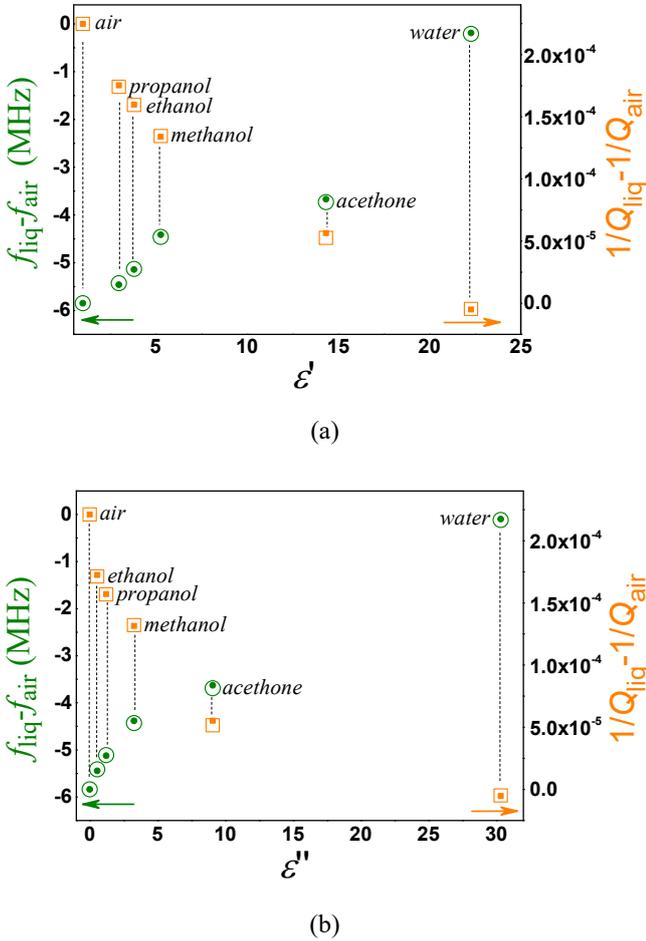

(a)

(b)

Fig. 3. Resonance frequency shift (circles) and changes in inverse $Q$-factors (squares) as a function of: (a) real and (b) imaginary parts of permittivity. The solid points present experimental data, open points calculated ones and dashed lines link data corresponding to the certain liquid.

The solid points present experimental data while open points calculated ones. Dashed lines link data points (frequency shift and changes in inverse quality factor) corresponding to the certain liquid. The frequency shift $f_{liq} - f_{air}$ increases with increasing of real part of liquid permittivity and becomes almost linear for points with permittivity higher than 5. The dependence of changes in inverse quality factor $1/Q_{liquid} - 1/Q_{air}$ on imaginary part of liquid permittivity is also almost linear for liquids with relatively high permittivity. The numerical calculations were performed using corrected geometrical dimensions of the plastic layer with a microfluidic channel as was described in the previous section. The calculated data are in good agreement with experimental ones, which proves reliability of the modeling calculation with the geometrical dimensions obtained as a result of fitting procedure.

*C. Measurement of glucose in water solutions*

Solutions of glucose in water were chosen as well-known, important for insulin-related research and readily available biological solutions for resonator measurements. Aqueous solutions of glucose were prepared just before the measurements by dissolving glucose in water with mass concentrations of glucose equal to 0.5, 1, 2, 4, 6, 8, 10, 15 and 20%. The measurements were performed at a frequency of 35.5 GHz and with the temperature stabilized at 25ºC ± 0.01°C. The dependencies of the resonant frequency shift $\Delta f = f_{solution} - f_{water}$ and changes in inverse $Q$-factors $\Delta(1/Q) = 1/Q_{water} - 1/Q_{solution}$ of the resonator with a microfluidic channel filled with water and aqueous glucose solution on the concentration of glucose are shown in Fig. 4. The data show linear behavior as a function of glucose concentration.

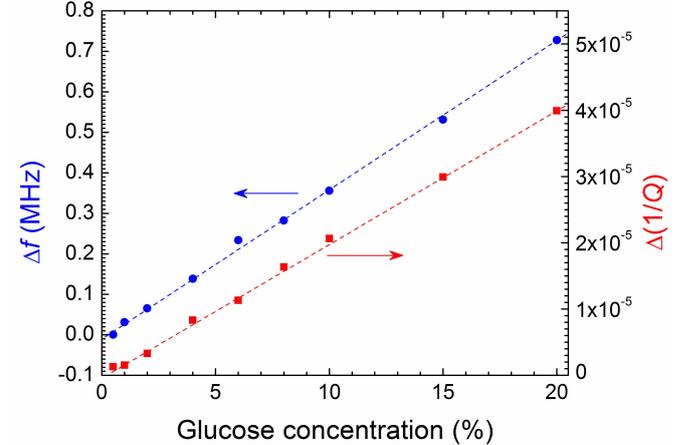

Fig. 4. Resonance frequency shift (circles) and changes in inverse $Q$-factors (squares) dependencies on concentration of glucose in aqueous solution.

*D. Solving the inverse problem for precise determination of liquid complex permittivity*

In the inverse electrodynamic problem, the real $\varepsilon'$ and imaginary $\varepsilon''$ parts of the permittivity of LUT are functions of both the resonant frequency shift and changes in the inverse $Q$-factor $\varepsilon' = f_1(\Delta f, \Delta(1/Q))$, $\varepsilon'' = f_2(\Delta f, \Delta(1/Q))$. It should be noted that the values measured in the direct experiment are the eigenfrequencies and the magnitudes of the $Q$-factor and these values depend on $\varepsilon'$ and $\varepsilon''$. In this case, it is not possible to obtain the absolute value of permittivity for LUT by calibration using the known liquids. For this purpose, simulation by COMSOL Multiphysics can be used as will be shown below. As was shown previously [15], the modeling data is in precisely good agreement with the experimental data. Therefore, in order to solve the inverse problem, the calculations of the resonant frequency shift and changes in inverse $Q$-factor were performed for the microfluidic channel filled with a model substance with variable dielectric properties. The real part of the permittivity was varied in the interval of $\varepsilon'$=22.277-17.277 at a certain value of $\varepsilon''$ and the imaginary part of the permittivity was varied in the interval of $\varepsilon''$=30.337-22.337 at a certain value of $\varepsilon'$ with a definite step. As a result, the nomographic chart of the calculation results was constructed (Fig. 5). The experimental data (solid points) on measurements of the aqueous solution of glucose are also shown in Fig. 5. The nomogram shows that changing the imaginary part of the permittivity has a stronger influence on the resonant frequency shift than on changes in the inverse $Q$-factor and, vice versa, the real part of the permittivity has a stronger influence on changes in the inverse $Q$-factor. Fig. 5 allows the correspondence of the calculated complex permittivity net with experimentally obtained data (red points) for resonant frequency shift and changes in inverse $Q$-factor. For example, for 20% glucose solution point the real part of permittivity was obtained as 18.1 and imaginary part – as 23.0 (showed by dashed lines). Using such a nomogram allows us to determine the absolute value of the complex permittivity of the liquid under test.

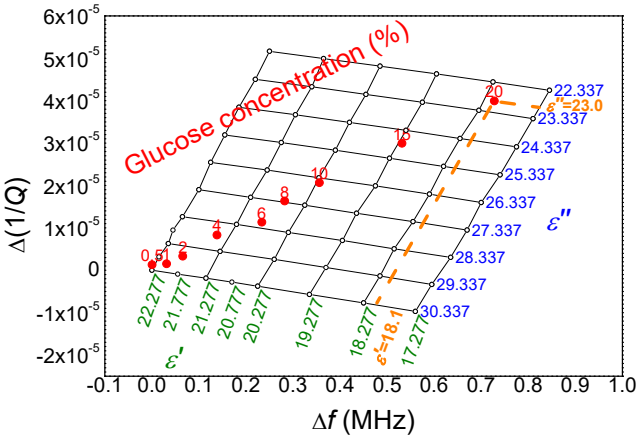

Fig. 5. Nomogram chart. Calculated net for the resonant frequency shift and changes in the inverse quality factor for a microfluidic channel filled with substances with varying real and imaginary parts of the permittivity (see text for details). Red points show the experimentally obtained data for resonant frequency shift and changes in inverse $Q$-factor as functions of glucose concentration in water solution.

## IV. PERMITTIVITY MEASUREMENT RESULTS

### A. Absolute values of complex permittivity of glucose solution

Fig. 6 shows the real and imaginary parts of permittivity dependencies on the concentration of glucose in a water solution obtained using the data from Fig. 5. Both dependencies display linear behavior as a function of glucose concentration.

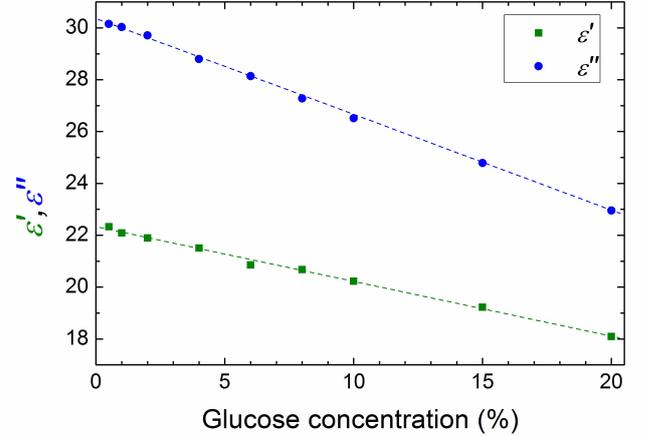

Fig. 6. Real (squares) and imaginary (circles) parts of permittivity dependencies on concentration of glucose in aqueous solutions.

The obtained data are in good agreement with those obtained in [17]. The difference between values of permittivity of the glucose solutions obtained using our technique and the values from [17] do not exceed 3%. If the temperature dependence of the relative permittivity of a glucose solution is taken into account, then this deviation can be neglected. This fact proves that the developed technique is precise and reliable.

### B. Investigation of albumins and protein aqueous solutions

The high sensitivity of WGM resonator technique promises possibility to study biochemical and biological liquids. The albumin bovine serum, lactalbumin and cytochrome C solutions in water were investigated. The concentration varies from 0.01 up to 3.0 mmol/l. The real and imaginary parts of liquid permittivity's dependencies on concentration of the mentioned biological substances in aqueous solutions are shown in Fig. 7.

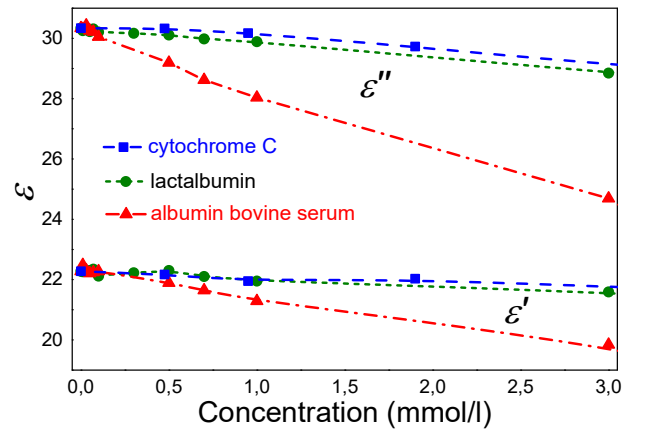

Fig. 7. Real and imaginary parts of permittivity dependencies on concentration of albumin bovine serum (triangles and dash-dot line), lactalbumin (circles and short-dashed line) and cytochrome C (squares and dashed line) in aqueous solutions.

The relative accuracy of the complex permittivity values of the glucose water solutions obtained using described technique is about 1.4% for real part and 0.7% for the imaginary part. These values are evaluated as $\delta\varepsilon'/\varepsilon'$ and $\delta\varepsilon''/\varepsilon''$, respectively. Here $\delta\varepsilon'$ and $\delta\varepsilon''$ are the most probable errors for corresponding permittivity parts [12,14].

The measurements show not only possibility of studying the aqueous solutions of different compounds with the small concentration changes, but also show possibility of quantitative monitoring of different LUT behavior using proposed novel technique.

## V. Conclusion

A novel approach for obtaining the absolute values of complex permittivity of a liquid under test in the microwave range using the WGM resonator with a microfluidic channel on a top of the resonator has been proposed and successfully verified. The permittivity values of glucose aqueous solutions obtained using the developed procedures are in good agreement with literature data. The complex permittivities of albumin bovine serum, lactalbumin and cytochrome C aqueous solutions of various concentrations were obtained on the basis of measured data of the WGM resonator technique. It was shown that the presented measurement system allows measuring small concentration changes of liquid solutions in sub-microliter/nanoliter volumes. The approach enables the contactless determination of complex permittivity for small amounts of biological liquids in the microwave range with high measurement accuracy.

### Acknowledgments

The work was performed in the framework of the Cooperation Agreement between Forschungszentrum Jülich GmbH, Federal Republic of Germany, and O.Usikov Institute for Radiophysics and Electronics, National Academy of Science of Ukraine.